
  \input miniltx
  \def\Gin@driver{pdftex.def}
  \input color.sty
  \input graphicx.sty
  \resetatcatcode

%
%

%
%
%
%

\def\Serif{cmr}
\def\SerifBold{cmbx}
\def\SerifItalics{cmti}
\def\SerifSlanted{cmsl}
\def\SerifBoldItalics{cmbxti}
\def\SansSerif{cmss}
\def\SansSerifBold{cmssbx}
\def\SansSerifItalics{cmssi}
\def\SansSerifSlanted{cmssi}
\def\Math{cmmi}
\def\Symbols{cmsy}
\def\MathBold{cmmib}
\def\MoreSymbols{cmex}
\def\Typewriter{cmtt}
\def\Gothic{eufm}
\def\Double{msbm}
\def\Relazioni{msam}

= 			\Serif10 			at 5pt
= 		\SerifBold10 		at 5pt
= 	\SerifItalics10 	at 5pt
=		\SerifSlanted10 	at 5pt
=	\SerifBoldItalics10	at 5pt
= 		\SansSerif10 		at 5pt
=	\SansSerifBold10	at 5pt
=	\SansSerifItalics10	at 5pt
=	\SansSerifSlanted10	at 5pt
=				\Math10				at 5pt
=			\MathBold10			at 5pt
=			\Symbols10			at 5pt
=		\MoreSymbols10		at 5pt
=		\Typewriter10		at 5pt
=			\Gothic10			at 5pt
=			\Double10			at 5pt

= 			\Serif10 			at 7pt
= 		\SerifBold10 		at 7pt
= 	\SerifItalics10 	at 7pt
=	\SerifSlanted10 	at 7pt
=\SerifBoldItalics10	at 7pt
= 		\SansSerif10 		at 7pt
= 	\SansSerifBold10 	at 7pt
=\SansSerifItalics10	at 7pt
=\SansSerifSlanted10	at 7pt
=			\Math10				at 7pt
=		\MathBold10			at 7pt
=			\Symbols10			at 7pt
=		\MoreSymbols10		at 7pt
=		\Typewriter10		at 7pt
=			\Gothic10			at 7pt
=			\Double10			at 7pt

= 			\Serif10 			at 8pt
= 		\SerifBold10 		at 8pt
= 	\SerifItalics10 	at 8pt
=	\SerifSlanted10 	at 8pt
=\SerifBoldItalics10	at 8pt
= 		\SansSerif10 		at 8pt
= 	\SansSerifBold10 	at 8pt
=\SansSerifItalics10 at 8pt
=\SansSerifSlanted10 at 8pt
=			\Math10				at 8pt
=		\MathBold10			at 8pt
=			\Symbols10			at 8pt
=		\MoreSymbols10		at 8pt
=		\Typewriter10		at 8pt
=			\Gothic10			at 8pt
=			\Double10			at 8pt

= 			\Serif10 			at 10pt
= 		\SerifBold10 		at 10pt
= 		\SerifItalics10 	at 10pt
=		\SerifSlanted10 	at 10pt
=	\SerifBoldItalics10	at 10pt
= 		\SansSerif10 		at 10pt
= 	\SansSerifBold10 	at 10pt
= 	\SansSerifItalics10 at 10pt
= 	\SansSerifSlanted10 at 10pt
=				\Math10				at 10pt
=			\MathBold10			at 10pt
=			\Symbols10			at 10pt
=		\MoreSymbols10		at 10pt
=		\Typewriter10		at 10pt
=			\Gothic10			at 10pt
=			\Double10			at 10pt
=			\Relazioni10			at 10pt

= 				\Serif10 			at 12pt
= 			\SerifBold10 		at 12pt
= 		\SerifItalics10 	at 12pt
=		\SerifSlanted10 	at 12pt
=	\SerifBoldItalics10	at 12pt
= 			\SansSerif10 		at 12pt
= 		\SansSerifBold10 	at 12pt
= 	\SansSerifItalics10 at 12pt
= 	\SansSerifSlanted10 at 12pt
=				\Math10				at 12pt
=			\MathBold10			at 12pt
=			\Symbols10			at 12pt
=		\MoreSymbols10		at 12pt
=			\Typewriter10		at 12pt
=				\Gothic10			at 12pt
=				\Double10			at 12pt

= 			\Serif10 			at 14pt
= 		\SerifBold10 		at 14pt
= 	\SerifItalics10 	at 14pt
=		\SerifSlanted10 	at 14pt
=	\SerifBoldItalics10	at 14pt
= 		\SansSerif10 		at 14pt
= 	\SansSerifBold10 	at 14pt
= \SansSerifSlanted10 at 14pt
= \SansSerifItalics10 at 14pt
=				\Math10				at 14pt
=			\MathBold10			at 14pt
=			\Symbols10			at 14pt
=		\MoreSymbols10		at 14pt
=		\Typewriter10		at 14pt
=			\Gothic10			at 14pt
=			\Double10			at 14pt

\def\NormalStyle{\parindent=5pt\parskip=3pt\normalbaselineskip=14pt%
\def\nt{\tenSerif}%
\def\rm{\fam0\tenSerif}%
\textfont0=\tenSerif\scriptfont0=\sevenSerif\scriptscriptfont0=\fiveSerif
\textfont1=\tenMath\scriptfont1=\sevenMath\scriptscriptfont1=\fiveMath
\textfont2=\tenSymbols\scriptfont2=\sevenSymbols\scriptscriptfont2=\fiveSymbols
\textfont3=\tenMoreSymbols\scriptfont3=\sevenMoreSymbols\scriptscriptfont3=\fiveMoreSymbols
\textfont\itfam=\tenSerifItalics\def\it{\fam\itfam\tenSerifItalics}%
\textfont\slfam=\tenSerifSlanted\def\sl{\fam\slfam\tenSerifSlanted}%
\textfont\ttfam=\tenTypewriter\def\tt{\fam\ttfam\tenTypewriter}%
\textfont\bffam=\tenSerifBold%
\def\bf{\fam\bffam\tenSerifBold}\scriptfont\bffam=\sevenSerifBold\scriptscriptfont\bffam=\fiveSerifBold%
\def\cal{\tenSymbols}%
\def\greekbold{\tenMathBold}%
\def\gothic{\tenGothic}%
\def\Bbb{\tenDouble}%
\def\LieFont{\tenSerifItalics}%
\nt\normalbaselines\baselineskip=14pt%
}

\def\TitleStyle{\parindent=0pt\parskip=0pt\normalbaselineskip=15pt%
\def\nt{\fourteenSansSerifBold}%
\def\rm{\fam0\fourteenSansSerifBold}%
\textfont0=\fourteenSansSerifBold\scriptfont0=\tenSansSerifBold\scriptscriptfont0=\eightSansSerifBold
\textfont1=\fourteenMath\scriptfont1=\tenMath\scriptscriptfont1=\eightMath
\textfont2=\fourteenSymbols\scriptfont2=\tenSymbols\scriptscriptfont2=\eightSymbols
\textfont3=\fourteenMoreSymbols\scriptfont3=\tenMoreSymbols\scriptscriptfont3=\eightMoreSymbols
\textfont\itfam=\fourteenSansSerifItalics\def\it{\fam\itfam\fourteenSansSerifItalics}%
\textfont\slfam=\fourteenSansSerifSlanted\def\sl{\fam\slfam\fourteenSerifSansSlanted}%
\textfont\ttfam=\fourteenTypewriter\def\tt{\fam\ttfam\fourteenTypewriter}%
\textfont\bffam=\fourteenSansSerif%
\def\bf{\fam\bffam\fourteenSansSerif}\scriptfont\bffam=\tenSansSerif\scriptscriptfont\bffam=\eightSansSerif%
\def\cal{\fourteenSymbols}%
\def\greekbold{\fourteenMathBold}%
\def\gothic{\fourteenGothic}%
\def\Bbb{\fourteenDouble}%
\def\LieFont{\fourteenSerifItalics}%
\nt\normalbaselines\baselineskip=15pt%
}

\def\PartStyle{\parindent=0pt\parskip=0pt\normalbaselineskip=15pt%
\def\nt{\fourteenSansSerifBold}%
\def\rm{\fam0\fourteenSansSerifBold}%
\textfont0=\fourteenSansSerifBold\scriptfont0=\tenSansSerifBold\scriptscriptfont0=\eightSansSerifBold
\textfont1=\fourteenMath\scriptfont1=\tenMath\scriptscriptfont1=\eightMath
\textfont2=\fourteenSymbols\scriptfont2=\tenSymbols\scriptscriptfont2=\eightSymbols
\textfont3=\fourteenMoreSymbols\scriptfont3=\tenMoreSymbols\scriptscriptfont3=\eightMoreSymbols
\textfont\itfam=\fourteenSansSerifItalics\def\it{\fam\itfam\fourteenSansSerifItalics}%
\textfont\slfam=\fourteenSansSerifSlanted\def\sl{\fam\slfam\fourteenSerifSansSlanted}%
\textfont\ttfam=\fourteenTypewriter\def\tt{\fam\ttfam\fourteenTypewriter}%
\textfont\bffam=\fourteenSansSerif%
\def\bf{\fam\bffam\fourteenSansSerif}\scriptfont\bffam=\tenSansSerif\scriptscriptfont\bffam=\eightSansSerif%
\def\cal{\fourteenSymbols}%
\def\greekbold{\fourteenMathBold}%
\def\gothic{\fourteenGothic}%
\def\Bbb{\fourteenDouble}%
\def\LieFont{\fourteenSerifItalics}%
\nt\normalbaselines\baselineskip=15pt%
}

\def\ChapterStyle{\parindent=0pt\parskip=0pt\normalbaselineskip=15pt%
\def\nt{\fourteenSansSerifBold}%
\def\rm{\fam0\fourteenSansSerifBold}%
\textfont0=\fourteenSansSerifBold\scriptfont0=\tenSansSerifBold\scriptscriptfont0=\eightSansSerifBold
\textfont1=\fourteenMath\scriptfont1=\tenMath\scriptscriptfont1=\eightMath
\textfont2=\fourteenSymbols\scriptfont2=\tenSymbols\scriptscriptfont2=\eightSymbols
\textfont3=\fourteenMoreSymbols\scriptfont3=\tenMoreSymbols\scriptscriptfont3=\eightMoreSymbols
\textfont\itfam=\fourteenSansSerifItalics\def\it{\fam\itfam\fourteenSansSerifItalics}%
\textfont\slfam=\fourteenSansSerifSlanted\def\sl{\fam\slfam\fourteenSerifSansSlanted}%
\textfont\ttfam=\fourteenTypewriter\def\tt{\fam\ttfam\fourteenTypewriter}%
\textfont\bffam=\fourteenSansSerif%
\def\bf{\fam\bffam\fourteenSansSerif}\scriptfont\bffam=\tenSansSerif\scriptscriptfont\bffam=\eightSansSerif%
\def\cal{\fourteenSymbols}%
\def\greekbold{\fourteenMathBold}%
\def\gothic{\fourteenGothic}%
\def\Bbb{\fourteenDouble}%
\def\LieFont{\fourteenSerifItalics}%
\nt\normalbaselines\baselineskip=15pt%
}

\def\SectionStyle{\parindent=0pt\parskip=0pt\normalbaselineskip=13pt%
\def\nt{\twelveSansSerifBold}%
\def\rm{\fam0\twelveSansSerifBold}%
\textfont0=\twelveSansSerifBold\scriptfont0=\eightSansSerifBold\scriptscriptfont0=\eightSansSerifBold
\textfont1=\twelveMath\scriptfont1=\eightMath\scriptscriptfont1=\eightMath
\textfont2=\twelveSymbols\scriptfont2=\eightSymbols\scriptscriptfont2=\eightSymbols
\textfont3=\twelveMoreSymbols\scriptfont3=\eightMoreSymbols\scriptscriptfont3=\eightMoreSymbols
\textfont\itfam=\twelveSansSerifItalics\def\it{\fam\itfam\twelveSansSerifItalics}%
\textfont\slfam=\twelveSansSerifSlanted\def\sl{\fam\slfam\twelveSerifSansSlanted}%
\textfont\ttfam=\twelveTypewriter\def\tt{\fam\ttfam\twelveTypewriter}%
\textfont\bffam=\twelveSansSerif%
\def\bf{\fam\bffam\twelveSansSerif}\scriptfont\bffam=\eightSansSerif\scriptscriptfont\bffam=\eightSansSerif%
\def\cal{\twelveSymbols}%
\def\bg{\twelveMathBold}%
\def\gothic{\twelveGothic}%
\def\Bbb{\twelveDouble}%
\def\LieFont{\twelveSerifItalics}%
\nt\normalbaselines\baselineskip=13pt%
}

\def\SubSectionStyle{\parindent=0pt\parskip=0pt\normalbaselineskip=13pt%
\def\nt{\twelveSansSerifItalics}%
\def\rm{\fam0\twelveSansSerifItalics}%
\textfont0=\twelveSansSerifItalics\scriptfont0=\eightSansSerifItalics\scriptscriptfont0=\eightSansSerifItalics%
\textfont1=\twelveMath\scriptfont1=\eightMath\scriptscriptfont1=\eightMath%
\textfont2=\twelveSymbols\scriptfont2=\eightSymbols\scriptscriptfont2=\eightSymbols%
\textfont3=\twelveMoreSymbols\scriptfont3=\eightMoreSymbols\scriptscriptfont3=\eightMoreSymbols%
\textfont\itfam=\twelveSansSerif\def\it{\fam\itfam\twelveSansSerif}%
\textfont\slfam=\twelveSansSerifSlanted\def\sl{\fam\slfam\twelveSerifSansSlanted}%
\textfont\ttfam=\twelveTypewriter\def\tt{\fam\ttfam\twelveTypewriter}%
\textfont\bffam=\twelveSansSerifBold%
\def\bf{\fam\bffam\twelveSansSerifBold}\scriptfont\bffam=\eightSansSerifBold\scriptscriptfont\bffam=\eightSansSerifBold%
\def\cal{\twelveSymbols}%
\def\greekbold{\twelveMathBold}%
\def\gothic{\twelveGothic}%
\def\Bbb{\twelveDouble}%
\def\LieFont{\twelveSerifItalics}%
\nt\normalbaselines\baselineskip=13pt%
}

\def\AuthorStyle{\parindent=0pt\parskip=0pt\normalbaselineskip=14pt%
\def\nt{\tenSerif}%
\def\rm{\fam0\tenSerif}%
\textfont0=\tenSerif\scriptfont0=\sevenSerif\scriptscriptfont0=\fiveSerif
\textfont1=\tenMath\scriptfont1=\sevenMath\scriptscriptfont1=\fiveMath
\textfont2=\tenSymbols\scriptfont2=\sevenSymbols\scriptscriptfont2=\fiveSymbols
\textfont3=\tenMoreSymbols\scriptfont3=\sevenMoreSymbols\scriptscriptfont3=\fiveMoreSymbols
\textfont\itfam=\tenSerifItalics\def\it{\fam\itfam\tenSerifItalics}%
\textfont\slfam=\tenSerifSlanted\def\sl{\fam\slfam\tenSerifSlanted}%
\textfont\ttfam=\tenTypewriter\def\tt{\fam\ttfam\tenTypewriter}%
\textfont\bffam=\tenSerifBold%
\def\bf{\fam\bffam\tenSerifBold}\scriptfont\bffam=\sevenSerifBold\scriptscriptfont\bffam=\fiveSerifBold%
\def\cal{\tenSymbols}%
\def\greekbold{\tenMathBold}%
\def\gothic{\tenGothic}%
\def\Bbb{\tenDouble}%
\def\LieFont{\tenSerifItalics}%
\nt\normalbaselines\baselineskip=14pt%
}

\def\AddressStyle{\parindent=0pt\parskip=0pt\normalbaselineskip=14pt%
\def\nt{\eightSerif}%
\def\rm{\fam0\eightSerif}%
\textfont0=\eightSerif\scriptfont0=\sevenSerif\scriptscriptfont0=\fiveSerif
\textfont1=\eightMath\scriptfont1=\sevenMath\scriptscriptfont1=\fiveMath
\textfont2=\eightSymbols\scriptfont2=\sevenSymbols\scriptscriptfont2=\fiveSymbols
\textfont3=\eightMoreSymbols\scriptfont3=\sevenMoreSymbols\scriptscriptfont3=\fiveMoreSymbols
\textfont\itfam=\eightSerifItalics\def\it{\fam\itfam\eightSerifItalics}%
\textfont\slfam=\eightSerifSlanted\def\sl{\fam\slfam\eightSerifSlanted}%
\textfont\ttfam=\eightTypewriter\def\tt{\fam\ttfam\eightTypewriter}%
\textfont\bffam=\eightSerifBold%
\def\bf{\fam\bffam\eightSerifBold}\scriptfont\bffam=\sevenSerifBold\scriptscriptfont\bffam=\fiveSerifBold%
\def\cal{\eightSymbols}%
\def\greekbold{\eightMathBold}%
\def\gothic{\eightGothic}%
\def\Bbb{\eightDouble}%
\def\LieFont{\eightSerifItalics}%
\nt\normalbaselines\baselineskip=14pt%
}

\def\AbstractStyle{\parindent=0pt\parskip=0pt\normalbaselineskip=12pt%
\def\nt{\eightSerif}%
\def\rm{\fam0\eightSerif}%
\textfont0=\eightSerif\scriptfont0=\sevenSerif\scriptscriptfont0=\fiveSerif
\textfont1=\eightMath\scriptfont1=\sevenMath\scriptscriptfont1=\fiveMath
\textfont2=\eightSymbols\scriptfont2=\sevenSymbols\scriptscriptfont2=\fiveSymbols
\textfont3=\eightMoreSymbols\scriptfont3=\sevenMoreSymbols\scriptscriptfont3=\fiveMoreSymbols
\textfont\itfam=\eightSerifItalics\def\it{\fam\itfam\eightSerifItalics}%
\textfont\slfam=\eightSerifSlanted\def\sl{\fam\slfam\eightSerifSlanted}%
\textfont\ttfam=\eightTypewriter\def\tt{\fam\ttfam\eightTypewriter}%
\textfont\bffam=\eightSerifBold%
\def\bf{\fam\bffam\eightSerifBold}\scriptfont\bffam=\sevenSerifBold\scriptscriptfont\bffam=\fiveSerifBold%
\def\cal{\eightSymbols}%
\def\greekbold{\eightMathBold}%
\def\gothic{\eightGothic}%
\def\Bbb{\eightDouble}%
\def\LieFont{\eightSerifItalics}%
\nt\normalbaselines\baselineskip=12pt%
}

\def\RefsStyle{\parindent=0pt\parskip=0pt%
\def\nt{\eightSerif}%
\def\rm{\fam0\eightSerif}%
\textfont0=\eightSerif\scriptfont0=\sevenSerif\scriptscriptfont0=\fiveSerif
\textfont1=\eightMath\scriptfont1=\sevenMath\scriptscriptfont1=\fiveMath
\textfont2=\eightSymbols\scriptfont2=\sevenSymbols\scriptscriptfont2=\fiveSymbols
\textfont3=\eightMoreSymbols\scriptfont3=\sevenMoreSymbols\scriptscriptfont3=\fiveMoreSymbols
\textfont\itfam=\eightSerifItalics\def\it{\fam\itfam\eightSerifItalics}%
\textfont\slfam=\eightSerifSlanted\def\sl{\fam\slfam\eightSerifSlanted}%
\textfont\ttfam=\eightTypewriter\def\tt{\fam\ttfam\eightTypewriter}%
\textfont\bffam=\eightSerifBold%
\def\bf{\fam\bffam\eightSerifBold}\scriptfont\bffam=\sevenSerifBold\scriptscriptfont\bffam=\fiveSerifBold%
\def\cal{\eightSymbols}%
\def\greekbold{\eightMathBold}%
\def\gothic{\eightGothic}%
\def\Bbb{\eightDouble}%
\def\LieFont{\eightSerifItalics}%
\nt\normalbaselines\baselineskip=10pt%
}



%
%


\def\ModeYes{yes}
\def\ModeNo{no}

\def\ModeUndef{undefined}


\def\nx{\noexpand}
\def\ni{\noindent}
\def\newpage{\vfill\eject}

\def\ss{\vskip 5pt}
\def\ms{\vskip 10pt}
\def\bs{\vskip 20pt}

 \def\,{\mskip\thinmuskip}
 \def\!{\mskip-\thinmuskip}
 \def\>{\mskip\medmuskip}
 \def\;{\mskip\thickmuskip}

%
% ... 
%

\def\refsModePost{post}
\def\refsModeAuto{auto}

\def\dbRefsSatusModeOk{ok}
\def\dbRefsSatusModeError{error}
\def\dbRefsSatusModeWarning{warning}


\newcount\BNUM
\BNUM=0

\def\refs{}

\def\SetModePost{\xdef\refsMode{\refsModePost}}			
\SetModePost

\def\dbRefsStatusOk{%
	\xdef\dbRefsStatus{\dbRefsSatusModeOk}%
	\xdef\dbRefsError{\ModeNo}%
	\xdef\dbRefsWarning{\ModeNo}%
	\xdef\dbRefsInfo{\ModeNo}%
}

\def\dbRefs{%
}

\def\dbRefsGet#1{%
	\xdef\found{N}\xdef\ikey{#1}\dbRefsStatusOk%
	\xdef\key{\ModeUndef}\xdef\tag{\ModeUndef}\xdef\tail{\ModeUndef}%
	\dbRefs%
}

\def\NextRefsTag{%
	\global\advance\BNUM by 1%
}
\def\ShowTag#1{{\bf [#1]}}

\def\dbRefsInsert#1#2{%
\dbRefsGet{#1}%
\if\found Y %
   \xdef\dbRefsStatus{\dbRefsSatusModeWarning}%
   \xdef\dbRefsWarning{record is already there}%
   \xdef\dbRefsInfo{record not inserted}%
\else%
   \toks2=\expandafter{\dbRefs}%
   \ifx\refsMode\refsModeAuto \NextRefsTag
    \xdef\dbRefs{%
   	\the\toks2 \nx\xdef\nx\dbx{#1}%
	\nx\ifx\nx\ikey %
		\nx\dbx\nx\xdef\nx\found{Y}%
		\nx\xdef\nx\key{#1}%
		\nx\xdef\nx\tag{\the\BNUM}%
		\nx\xdef\nx\tail{#2}%
	\nx\fi}%
	\global\xdef\refs{\refs \ss\ni[\the\BNUM]\ #2\par}
   \fi%
   \ifx\refsMode\refsModePost 
    \xdef\dbRefs{%
   	\the\toks2 \nx\xdef\nx\dbx{#1}%
	\nx\ifx\nx\ikey %
		\nx\dbx\nx\xdef\nx\found{Y}%
		\nx\xdef\nx\key{#1}%
		\nx\xdef\nx\tag{\ModeUndef}%
		\nx\xdef\nx\tail{#2}%
	\nx\fi}%
   \fi%
\fi%
}

\def\dbRefsEdit#1#2#3{\dbRefsGet{#1}%
\if\found N 
   \xdef\dbRefsStatus{\dbRefsSatusModeError}%
   \xdef\dbRefsError{record is not there}%
   \xdef\dbRefsInfo{record not edited}%
\else%
   \toks2=\expandafter{\dbRefs}%
   \xdef\dbRefs{\the\toks2%
   \nx\xdef\nx\dbx{#1}%
   \nx\ifx\nx\ikey\nx\dbx %
	\nx\xdef\nx\found{Y}%
	\nx\xdef\nx\key{#1}%
	\nx\xdef\nx\tag{#2}%
	\nx\xdef\nx\tail{#3}%
   \nx\fi}%
\fi%
}

\def\bib#1#2{\RefsStyle\dbRefsInsert{#1}{#2}%
	\ifx\dbRefsStatus\dbRefsSatusModeWarning %
		\message{^^J}%
		\message{WARNING: Reference [#1] is doubled.^^J}%
	\fi%
}

\def\ref#1{\dbRefsGet{#1}%
\ifx\found N %
  \message{^^J}%
  \message{ERROR: Reference [#1] unknown.^^J}%
  \ShowTag{??}%
\else%
	\ifx\tag\ModeUndef \NextRefsTag%
		\dbRefsEdit{#1}{\the\BNUM}{\tail}%
		\dbRefsGet{#1}%
		\global\xdef\refs{\refs \ss\ni [\tag]\ \tail\par}
	\fi
	\ShowTag{\tag}%
\fi%
}

\def\ShowBiblio{\ms\Ensure{\SectionEnsure}%
{\SectionStyle\ni References}%
{\RefsStyle\refs}%
}

\newcount\CHANGES
\CHANGES=0
\def\AuxFile{7}
\def\PreventDoubleOn{\xdef\PreventDoubleLabel{\ModeYes}}

\PreventDoubleOn

\def\StoreLabel#1#2{\xdef\itag{#2}
 \ifx\PreModeStatus\ModeNo %
   \message{^^J}%
   \errmessage{You can't use Check without starting with OpenPreMode (and finishing with ClosePreMode)^^J}%
 \else%
   \immediate\write\AuxFile{\nx\dbLabelPreInsert{#1}{\itag}}%
   \dbLabelGet{#1}%
   \ifx\itag\tag %
   \else%
	\global\advance\CHANGES by 1%
 	\xdef\itag{(?.??)}%
    \fi%
   \fi%
}

\def\PreModeStatus{\ModeNo}

\def\ClosePreMode{\immediate\closeout\AuxFile%
  \ifnum\CHANGES=0%
	\message{^^J}%
	\message{**********************************^^J}%
	\message{**  NO CHANGES TO THE AuxFile  **^^J}%
	\message{**********************************^^J}%
 \else%
	\message{^^J}%
	\message{**************************************************^^J}%
	\message{**  PLAEASE TYPESET IT AGAIN (\the\CHANGES)  **^^J}%
    \errmessage{**************************************************^^ J}%
  \fi%
  \edef\PreModeStatus{\ModeNo}%
}

\def\dbLabelSatusModeOk{ok}

\def\dbLabelSatusModeWarning{warning}

\def\dbLabelStatusOk{%
	\xdef\dbLabelStatus{\dbLabelSatusModeOk}%
	\xdef\dbLabelError{\ModeNo}%
	\xdef\dbLabelWarning{\ModeNo}%
	\xdef\dbLabelInfo{\ModeNo}%
}

\def\dbLabel{%
}

\def\dbLabelGet#1{%
	\xdef\found{N}\xdef\ikey{#1}\dbLabelStatusOk%
	\xdef\key{\ModeUndef}\xdef\tag{\ModeUndef}\xdef\pre{\ModeUndef}%
	\dbLabel%
}

\def\ShowLabel#1{%
 \dbLabelGet{#1}%
 \ifx\tag \ModeUndef %
 	\global\advance\CHANGES by 1%
 	(?.??)%
 \else%
 	\tag%
 \fi%
}

\def\dbLabelPreInsert#1#2{\dbLabelGet{#1}%
\if\found Y %
  \xdef\dbLabelStatus{\dbLabelSatusModeWarning}%
   \xdef\dbLabelWarning{Label is already there}%
   \xdef\dbLabelInfo{Label not inserted}%
   \message{^^J}%
   \errmessage{Double pre definition of label [#1]^^J}%
\else%
   \toks2=\expandafter{\dbLabel}%
    \xdef\dbLabel{%
   	\the\toks2 \nx\xdef\nx\dbx{#1}%
	\nx\ifx\nx\ikey %
		\nx\dbx\nx\xdef\nx\found{Y}%
		\nx\xdef\nx\key{#1}%
		\nx\xdef\nx\tag{#2}%
		\nx\xdef\nx\pre{\ModeYes}%
	\nx\fi}%
\fi%
}

\def\dbLabelInsert#1#2{\dbLabelGet{#1}%
\xdef\itag{#2}%
\dbLabelGet{#1}%
\if\found Y %
	\ifx\tag\itag %
	\else%
	   \ifx\PreventDoubleLabel\ModeYes %
		\message{^^J}%
		\errmessage{Double definition of label [#1]^^J}%
	   \else%
		\message{^^J}%
		\message{Double definition of label [#1]^^J}%
	   \fi%
	\fi%
   \xdef\dbLabelStatus{\dbLabelSatusModeWarning}%
   \xdef\dbLabelWarning{Label is already there}%
   \xdef\dbLabelInfo{Label not inserted}%
\else%
   \toks2=\expandafter{\dbLabel}%
    \xdef\dbLabel{%
   	\the\toks2 \nx\xdef\nx\dbx{#1}%
	\nx\ifx\nx\ikey %
		\nx\dbx\nx\xdef\nx\found{Y}%
		\nx\xdef\nx\key{#1}%
		\nx\xdef\nx\tag{#2}%
		\nx\xdef\nx\pre{\ModeNo}%
	\nx\fi}%
\fi%
}


\newcount\PART
\newcount\CHAPTER
\newcount\SECTION
\newcount\SUBSECTION
\newcount\FNUMBER

\PART=0
\CHAPTER=0
\SECTION=0
\SUBSECTION=0	
\FNUMBER=0

\def\LastPart{\ModeUndef}
\def\LastChapter{\ModeUndef}
\def\LastSection{\ModeUndef}
\def\LastSubSection{\ModeUndef}
\def\LastClaim{\ModeUndef}
\def\Last{\ModeUndef}

\newdimen\TOBOTTOM
\newdimen\LIMIT

\def\Ensure#1{\ \par\ \immediate\LIMIT=#1\immediate\TOBOTTOM=\the\pagegoal\advance\TOBOTTOM by -\pagetotal%
\ifdim\TOBOTTOM<\LIMIT\newpage \else%
\vskip-\parskip\vskip-\parskip\vskip-\baselineskip\fi}

\def\PartLabel{\the\PART}
\def\NewPart#1{\global\advance\PART by 1%
         \bs\ni{\PartStyle  Part \PartLabel:}
         \bs\ni{\PartStyle #1}\newpage%
         \CHAPTER=0\SECTION=0\SUBSECTION=0\FNUMBER=0%
         \gdef\Left{#1}%
         \global\edef\Last{\PartLabel}%
         \global\edef\LastPart{\PartLabel}%
         \global\edef\LastChapter{\ModeUndef}%
         \global\edef\LastSection{\ModeUndef}%
         \global\edef\LastSubSection{\ModeUndef}%
         \global\edef\LastClaim{\ModeUndef}}
\def\ChapterLabel{\the\CHAPTER}
\def\NewChapter#1{\global\advance\CHAPTER by 1%
         \bs\ni{\ChapterStyle  Chapter \ChapterLabel: #1}\ms%
         \SECTION=0\SUBSECTION=0\FNUMBER=0%
         \gdef\Left{#1}%
         \global\edef\Last{\ChapterLabel}%
         \global\edef\LastChapter{\ChapterLabel}%
         \global\edef\LastSection{\ModeUndef}%
         \global\edef\LastSubSection{\ModeUndef}%
         \global\edef\LastClaim{\ModeUndef}}
\def\SectionEnsure{3cm}
\def\NewSection#1{\Ensure{\SectionEnsure}\gdef\SectionLabel{\the\SECTION}\global\advance\SECTION by 1%
         \ms\ni{\SectionStyle  \SectionLabel.\ #1}\ss%
         \SUBSECTION=0\FNUMBER=0%
         \gdef\Left{#1}%
         \global\edef\Last{\SectionLabel}%
         \global\edef\LastSection{\SectionLabel}%
         \global\edef\LastSubSection{\ModeUndef}%
         \global\edef\LastClaim{\ModeUndef}}
\def\NewAppendix#1#2{\Ensure{\SectionEnsure}\gdef\SectionLabel{#1}\global\advance\SECTION by 1%
         \bs\ni{\SectionStyle  Appendix \SectionLabel.\ #2}\ss%
         \SUBSECTION=0\FNUMBER=0%
         \gdef\Left{#2}%
         \global\edef\Last{\SectionLabel}%
         \global\edef\LastSection{\SectionLabel}%
         \global\edef\LastSubSection{\ModeUndef}%
         \global\edef\LastClaim{\ModeUndef}}
\def\Acknowledgements{\Ensure{\SectionEnsure}\gdef\SectionLabel{}%
         \ms\ni{\SectionStyle  Acknowledgments}\ss%
         \SECTION=0\SUBSECTION=0\FNUMBER=0%
         \gdef\Left{}%
         \global\edef\Last{\ModeUndef}%
         \global\edef\LastSection{\ModeUndef}%
         \global\edef\LastSubSection{\ModeUndef}%
         \global\edef\LastClaim{\ModeUndef}}
\def\SubSectionEnsure{2cm}
\def\SubSectionLabel{\ifnum\SECTION>0 \the\SECTION.\fi\the\SUBSECTION}
\def\NewSubSection#1{\Ensure{\SubSectionEnsure}\global\advance\SUBSECTION by 1%
         \ms\ni{\SubSectionStyle #1}\ss%
         \global\edef\Last{\SubSectionLabel}%
         \global\edef\LastSubSection{\SubSectionLabel}}
\def\SetNumberingModeN{\def\ClaimLabel{(\the\FNUMBER)}}
\def\SetNumberingModeSN{\def\ClaimLabel{(\ifnum\SECTION>0 \SectionLabel.\fi%
      \the\FNUMBER)}}
\def\SetNumberingModeCSN{\def\ClaimLabel{(\ifnum\CHAPTER>0 \the\CHAPTER.\fi%
      \ifnum\SECTION>0 \SectionLabel.\fi%
      \the\FNUMBER)}}

\def\NewClaim{\global\advance\FNUMBER by 1%
    \ClaimLabel%
    \global\edef\LastClaim{\ClaimLabel}%
    \global\edef\Last{\ClaimLabel}}

\def\HideLabels{\xdef\ShowLabelsMode{\ModeNo}}
\HideLabels

\def\fn{\eqno{\NewClaim}} 
\def\fl#1{%
\ifx\ShowLabelsMode\ModeYes%
 \eqno{{\buildrel{\hbox{\AbstractStyle[#1]}}\over{\hfill\NewClaim}}}%
\else%
 \eqno{\NewClaim}%
\fi%
\dbLabelInsert{#1}{\ClaimLabel}}
\def\fprel#1{\global\advance\FNUMBER by 1\StoreLabel{#1}{\ClaimLabel}%
\ifx\ShowLabelsMode\ModeYes%
\eqno{{\buildrel{\hbox{\AbstractStyle[#1]}}\over{\hfill.\itag}}}%
\else%
 \eqno{\itag}%
\fi%
}

\def\cl#1{\global\advance\FNUMBER by 1\dbLabelInsert{#1}{\ClaimLabel}%
\ifx\ShowLabelsMode\ModeYes%
${\buildrel{\hbox{\AbstractStyle[#1]}}\over{\hfill\ClaimLabel}}$%
\else%
  $\ClaimLabel$%
\fi%
}
\def\cprel#1{\global\advance\FNUMBER by 1\StoreLabel{#1}{\ClaimLabel}%
\ifx\ShowLabelsMode\ModeYes%
${\buildrel{\hbox{\AbstractStyle[#1]}}\over{\hfill.\itag}}$%
\else%
  $\itag$%
\fi%
}

\def\Note{\ms\leftskip 3cm\rightskip 1.5cm\AbstractStyle}
\def\endNote{\par\leftskip 2cm\rightskip 0cm\NormalStyle\ss}


\parindent=7pt
\leftskip=2cm
\newcount\SideIndent
\newcount\SideIndentTemp
\SideIndent=0
\newdimen\SectionIndent
\SectionIndent=-8pt

\def\sidebar{\vrule height15pt width.2pt }
\def\endcorner{\hbox{\hbox{\vrule height6pt width.2pt}\vbox to6pt{\vfill\hbox
to4pt{\leaders\hrule height0.2pt\hfill}}}}
\def\begincorner{\hbox{\hbox{\vrule height6pt width.2pt}\vbox to6pt{\hbox
to4pt{\leaders\hrule height0.2pt\hfill}}}}
\def\endbegincorner{\hbox{\vbox to15pt{\endcorner\vskip-6pt\begincorner\vfill}}}
\def\SideShow{\SideIndentTemp=\SideIndent \ifnum \SideIndentTemp>0 
\loop\sidebar\hskip 2pt \advance\SideIndentTemp by-1\ifnum \SideIndentTemp>1 \repeat\fi}

\def\BeginSection{{\vbadness 100000 \par\ni\hskip\SectionIndent%
\SideShow\vbox to 15pt{\vfill\begincorner}}\global\advance\SideIndent by1\vskip-10pt}

\def\EndSection{{\vbadness 100000 \par\ni\global\advance\SideIndent by-1%
\hskip\SectionIndent\SideShow\vbox to15pt{\endcorner\vfill}\vskip-10pt}}

\def\EndBeginSection{{\vbadness 100000\par\ni%
\global\advance\SideIndent by-1\hskip\SectionIndent\SideShow
\vbox to15pt{\vfill\endbegincorner}}%
\global\advance\SideIndent by1\vskip-10pt}

\def\ShowBeginCorners#1{%
\SideIndentTemp =#1 \advance\SideIndentTemp by-1%
\ifnum \SideIndentTemp>0 %
\vskip-15truept\hbox{\kern 2truept\vbox{\hbox{\begincorner}%
\ShowBeginCorners{\SideIndentTemp}\vskip-3truept}}%
\fi%
}

\def\ShowEndCorners#1{%
\SideIndentTemp =#1 \advance\SideIndentTemp by-1%
\ifnum \SideIndentTemp>0 %
\vskip-15truept\hbox{\kern 2truept\vbox{\hbox{\endcorner}%
\ShowEndCorners{\SideIndentTemp}\vskip 2truept}}%
\fi%
}

\def\BeginSections#1{{\vbadness 100000 \par\ni\hskip\SectionIndent%
\SideShow\vbox to 15pt{\vfill\ShowBeginCorners{#1}}}\global\advance\SideIndent by#1\vskip-10pt}

\def\EndSections#1{{\vbadness 100000 \par\ni\global\advance\SideIndent by-#1%
\hskip\SectionIndent\SideShow\vbox to15pt{\vskip15pt\ShowEndCorners{#1}\vfill}\vskip-10pt}}

\def\EndBeginSections#1#2{{\vbadness 100000\par\ni%
\global\advance\SideIndent by-#1%
\hbox{\hskip\SectionIndent\SideShow\kern-2pt%
\vbox to15pt{\vskip15pt\ShowEndCorners{#1}\vskip4pt\ShowBeginCorners{#2}}}}%
\global\advance\SideIndent by#2\vskip-10pt}




%
%


\def\al{\alpha}
\def\be{\beta}
\def\de{\delta}

\def\ep{\epsilon}

\def\te{\theta}
\def\la{\lambda}

\def\om{\omega}

\def\vp{\varphi}

\def\ka{\kappa}

\def\Ga{\Gamma}



 \def\gotG{{\hbox{\gothic G}}}
 \def\gotP{{\hbox{\gothic P}}}
 



\def\ip{\hbox to4pt{\leaders\hrule height0.3pt\hfill}\vbox to8pt{\leaders\vrule width0.3pt\vfill}\kern 2pt}
 
\def\del{\partial}
\def\na{\nabla}

\def\then{\Rightarrow}

%
%

\NormalStyle
\SetNumberingModeSN
\PreventDoubleOn

\long\def\title#1{\centerline{\TitleStyle\ni#1}}

\long\def\author#1{\ms\centerline{\AuthorStyle by {\it #1}}}

\long\def\address#1{\ss\centerline{\AddressStyle #1}\par}
\long\def\moreaddress#1{\centerline{\AddressStyle #1}\par}
\def\abstract{\ms\leftskip 3cm\rightskip .5cm\AbstractStyle{\bf \ni Abstract:}\ }
\def\endabstract{\par\leftskip 2cm\rightskip 0cm\NormalStyle\ss}

\SetNumberingModeSN

\def\nab#1{{\buildrel #1\over \na}}
\def\frac[#1/#2]{\hbox{$#1\over#2$}}
\def\Frac[#1/#2]{{#1\over#2}}
\def\({\left(}
\def\){\right)}
\def\[{\left[}
\def\]{\right]}
\def\^#1{{}^{#1}_{\>\cdot}}
\def\_#1{{}_{#1}^{\>\cdot}}
\def\Label=#1{{\buildrel {\hbox{\fiveSerif \ShowLabel{#1}}}\over =}}
\def\<{\kern -1pt}


\def\CollapseAllCNotes{\long\def\CNote##1{}}
\def\ExpandAllCNotes{\long\def\CNote##1{%
\BeginSection
	\Note%
 		##1%
	\endNote%
\EndSection%
}}
\ExpandAllCNotes
%
%
%
%


\def\red#1{\textcolor{red}{#1}}
\def\blue#1{\textcolor{blue}{#1}}

\def\frame#1{\vbox{\hrule\hbox{\vrule\vbox{\kern2pt\hbox{\kern2pt#1\kern2pt}\kern2pt}\vrule}\hrule\kern-4pt}}

\def\uline#1{\underline{#1}}

\def\Box to #1#2#3{\frame{\vtop{\hbox to #1{\hfill #2 \hfill}\hbox to #1{\hfill #3 \hfill}}}}


\bib{EPS}{J.Ehlers, F.A.E.Pirani, A.Schild, 
{\it The Geometry of Free Fall and Light Propagation},
in General Relativity, ed. L.OÕRaifeartaigh (Clarendon, Oxford, 1972). 
}

\bib{EPS1}
{M.Di Mauro, L. Fatibene, M.Ferraris, M.Francaviglia, 
{\it Further Extended Theories of Gravitation: Part I },
Int. J. Geom. Methods Mod. Phys. Volume: 7, Issue: 5 (2010), pp. 887-898; gr-qc/0911.2841}

\bib{EPS2}
{L. Fatibene, M.Ferraris, M.Francaviglia, S.Mercadante,
{\it Further Extended Theories of Gravitation: Part II},
Int. J. Geom. Methods Mod. Phys. Volume: 7, Issue: 5 (2010), pp. 899-906; gr-qc/0911.284}

\bib{ELQG}
{L. Fatibene, M. Ferraris, M. Francaviglia,
{\it Extended Loop Quantum Gravity},
CQG 27(18) 185016 (2010); arXiv:1003.1619}

\bib{MGaCou}
{L.Fatibene, M.Francaviglia, S. Mercadante,
{\it Matter Lagrangians Coupled with Connections}
Int. J. Geom. Methods Mod. Phys. Volume: 7, Issue: 5 (2010), 1185-1189; arXiv: 0911.2981}

\bib{catalogo}{ H.Stephani, D.Kramer, M.Mac Callum,
{\it Exact Solutions Of Einstein's Field Equations},
Cambridge University Press (2003) }

\bib{Perlick}{V. Perlick, 
{\it Characterization of standard clocks by means of light rays and freely falling particles}
General Relativity and Gravitation,  {\bf 19}(11) (1987) 1059-1073}

\bib{Faraoni}{T.P. Sotiriou, V. Faraoni,
{\it  $f (R)$  theories of gravity}, (2008); 
arXiv: 0805.1726v2
}

\bib{Capozziello}{S. Capozziello, M. De Laurentis, V. Faraoni
{\it A bird's eye view of $f(R)$-gravity}
(2009); arXiv:0909.4672 
}

\bib{Magnano}{G. Magnano, L.M. Sokolowski, 
{\it On Physical Equivalence between Nonlinear Gravity Theories}
Phys.Rev. D50 (1994) 5039-5059; gr-qc/9312008
}

\bib{S2}{T.P. Sotiriou,
{\it $f(R)$ gravity, torsion and non-metricity},
Class. Quant. Grav. 26 (2009) 152001; gr-qc/0904.2774}

\bib{S3}{T.P. Sotiriou,
{\it Modified Actions for Gravity: Theory and Phenomenology},
Ph.D. Thesis; gr-qc/0710.4438}

\bib{C1}{S. Capozziello, M. Francaviglia,
{\it Extended Theories of Gravity and their Cosmological and Astrophysical Applications},
Journal of General Relativity and Gravitation 40 (2-3), (2008) 357-420.}

\bib{C2}{S. Capozziello, M.F. De Laurentis, M. Francaviglia, S. Mercadante,
{\it From Dark Energy and Dark Matter to Dark Metric},
Foundations of Physics 39 (2009) 1161-1176
gr-qc/0805.3642v4}

\bib{C4}{S. Capozziello, M. De Laurentis, M. Francaviglia, S. Mercadante,
{\it First Order Extended Gravity and the Dark Side of the Universe Ð II: Matching Observational Data},
Proceedings of the Conference ``Univers Invisibile'', Paris June 29 Ð July 3, 2009 
Ð to appear in 2010}

\bib{Olmo}{Olmo Sigh}

\bib{Schouten}{
J.A.Schouten,
{\it Ricci-Calculus: An Introduction to Tensor Analysis and its Geometrical Applications},
Springer Verlag (1954)}

\bib{Bibliopolis}{M.Francaviglia,
{\it Relativistic theories},
Quaderni del CNR-GNFM (Italy, 1988)
}

\bib{Moon}{T.Y.Moon, J.Lee, P.Oh,
{\it Conformal invariance in Einstein-Cartan-Weyl space},
Mod. Phys. Lett. A25, 3129 (2010); arXiv:0912.0432[gr-qc]
}



\def\ubal{\underline{\al}\kern1pt}
\def\obal{\overline{\al}\kern1pt}

\def\ubR{\underline{R}\kern1pt}
\def\obR{\overline{R}\kern1pt}
\def\ubom{\underline{\om}\kern1pt}
\def\obxi{\overline{\xi}\kern1pt}
\def\ubu{\underline{u}\kern1pt}
\def\ube{\underline{e}\kern1pt}
\def\obe{\overline{e}\kern1pt}

\NormalStyle
\CollapseAllCNotes

\title{Weyl Geometries and Timelike Geodesics}

\author{L.Fatibene$^{a,b}$, M.Francaviglia$^{a,b}$}

\address{$^a$ Department of Mathematics, University of Torino (Italy)}
\moreaddress{$^b$ INFN - Iniziativa Specifica Na12}

\abstract
In view of Ehlers-Pirani-Schild formalism,  since 1972 Weyl geometries should be considered to be the most appropriate and complete framework to represent (relativistic) gravitational fields.
We shall here show that in any given Lorentzian spacetime $(M,g)$ that admits global timelike vector fields
any such vector field $u$ determines an essentially unique Weyl geometry $([g], \Ga)$ such that
 $u$ is $\Ga$-geodesic (i.e.~parallel with respect to $\Ga$).
\endabstract

\NewSection{Introduction}
We shall here present a result in Geometry which is motivated and finds important applications in relativistic theories of gravitation.
Let us recall first that in the early '70s Ehlers, Pirani and Schild (EPS) proposed (see \ref{EPS}) a set of axioms to fully determine 
the geometric structure of a spacetime $M$ able to describe the fundamental physical properties of any reasonable gravitational field, starting from the worldlines of free falling particles and lightrays;
see also \ref{EPS1}.

They showed that few reasonable and physically well-founded axioms uniquely determine a conformal structure
(i.e.~a class of conformally equivalent Lorentzian metrics) and a projective structure (i.e.~a class of projectively equivalent torsionless connections) in $M$.
The conformal structure $\gotG=[g]$ is uniquely characterized by the distribution of lightcones (which in fact are conformally invariant)
while the projective structure $\gotP=[\Ga]$ is uniquely defined by  the geodesic trajectories in spacetime; see \ref{Schouten}.

The conformal structure divides vectors into {\it $[g]$-timelike}, {\it $[g]$-lightlike} and {\it $[g]$-spacelike} vectors. Let us stress that this characterization just depends on the conformal class and it is independent of the choice of a representative.
Analogously, the projective structure $\gotP$ is associated to geodesic trajectories which are also called {\it $[\Ga]$-geodesics}
or simply {\it $\Ga$-geodesics}. Of course also a metric structure $g$ determines a connection (and a projective structure) 
through its Levi-Civita connection; the geodesic trajectories determined by the Levi-Civita connection are also called $g$-geodesics.

The conformal and projective structures have to be {\it compatible} in the sense that $[g]$-lightlike $\Ga$-geodesics
are also $g$-geodesics.

We shall define a pair $([g], [\Ga])$ of compatible conformal and projective structures in $M$ to be an {\it EPS structure on spacetime} $M$.
A {\it Weyl geometry} is a pair $([g], \Ga)$ formed by a conformal structure $[g]$ and a {\it single} torsionless connection $\Ga$
in the class $\gotP$ chosen so that there exists a covector $A$ such that
$$
\nab{\Ga} g = 2A\otimes g
\fn$$
Ehlers, Pirani and Schild showed that an EPS structure uniquely determines a Weyl geometry.

Locally this amounts to fix the connection as follows
$$
\Ga^\al_{\be\mu} = \{g\}^\al_{\be\mu} + \(g^{\al\ep} g_{\be\mu} -2\de^\al_{(\be}\de^\ep_{\mu)}\) A_\ep
\fl{EPSCompatibility}$$
where $\{g\}^\al_{\be\mu}$ are the Christoffel symbols of $g$ in any given chart. 

The effects of the gravitational field in this framework are described by two objects.
The propagation of lightrays is governed by the conformal structure while the motion of (massive) particles
is governed by the connection $\Ga$.  
In a sense, the gravitational field is a mix of these two objects.
One has therefore more kinematical freedom in fitting observational data relying on the extra degrees of freedom
contained in the covector $A$.

Since the time of EPS paper \ref{EPS} and especially since late '90s (when cosmological dynamics has been shown to present an
unexpected acceleration) a plethora of modified models of the Universe has been proposed to explain these observations.
In some cases modified models resort to adding matter sources to the gravitational field; this approach leads to argue 
that about 95\% of gravitational sources in our Universe are unknown  (or better they are known only through their
{\it observed gravitational effects} -- which are the reason that requires their introduction). 
Morever, most of these unknown components (about 70\%) have quite peculiar properties (namely this part produces a negative pressure
and it is known as {\it dark energy}).

Another approach leads instead to modify dynamics of the gravitational field in order to account for observations without resorting to  
{\it dark sources}.
A class of examples of such modified models is formed by so-called  $f(R)$-models
where the Lagrangian assumed to described gravity is any (non--linear) function of the scalar curvature of a spacetime $(M, g, \Ga)$;
 see \ref{Capozziello},\ref{Faraoni}, \ref{C1}, \ref{C2}, \ref{C4}, \ref{Moon}.
It has been shown that $f(R)$-models (in their metric-affine formulation) 
naturally lead to an EPS setting in which the gravitational field is in fact described by a Weyl geometry; see \ref{EPS1} and \ref{EPS2}.

After EPS a whole field of research has grown to generalize 
the physical interpretation of standard General Relativity to Weyl geometries; see \ref{Perlick} and references quoted therein.
EPS setting has also proved to be effective in studying models with matter interacting with the connection; \ref{MGaCou}, \ref{S2}, \ref{S3}, \ref{Olmo}.

\ms
Hereafter, in Section 2 we shall show that given a conformal structure $[g]$ over a spacetime $M$ then
a congruence of $[g]$-timelike curves (which can be interpreted as a the flow of all worldlines of a fluid)
determines an essentially unique EPS--compatible connection $\Ga$.

This provides a mechanism to determine a Weyl geometry $([g], \Ga)$ out of potentially observable quantities 
(the flow of the fluid).
It also provides a tool to fully exploit the new aforementioned degrees of freedom to disentangle Geometry from gravitational effects. 
The final aim of this study is the describe a generic fluid on a fixed {\it reference background $g$}.
In this framework the gravitational field is encoded into the covector $A$, still depending on the representative $g$ 
which one is free to fix in the conformal structure.

In Section $3$ we shall review what happens if another representative $\tilde g$ for the conformal structure is chosen.
We shall determine how the conformal rescaling acts on all objects involved in the formalism.

In Section $4$ we shall review the main assumptions made in Section $2$ where the Weyl connection is determined by a
congruence of timelike $\Ga$-geodesics. 
In Section 2 it is assumed that a fluid in a relativistic theory is described by a geodesic $[g]$-timelike field $u$ together with two scalar fields representing the pressure $p$ and the density $\rho$. This issue is relevant since in presence of pressure (i.e.~with internal forces) one could 
expect fluid particles to move along non-geodesic trajectories. 
We shall show that in some relevant cases the fluid can be nevertheless described in terms of a geodesic field also in presence of pressure.

\NewSection{Determining Weyl Geometry}

Let us consider a conformal structure $(M, [g])$ and any $[g]$-timelike vector field $u$.
Of course such a generic vector field has no reason to represent a $g$-geodesic field.
Hence in standard GR (in which the metric structure $g$ determines both lightrays and particle worldlines through its Levi-Civita connection $\Ga=\{g\}$ that in this case plays the role of the gravitational field) 
the situation is pretty rigid: once a metric structure has been fixed then relatively few curves are allowed 
to represent the motion of a fluid. Only a congruence of $g$-timelike $g$-geodesics is allowed.

We shall show hereafter that Weyl geometries are much less rigid. 
Once a conformal structure has been fixed then potentially any $[g]$-timelike vectorfield can be used to represent a fluid.
In fact one can use the extra degrees of freedom encoded by the covector $A$ in order to tune up the Weyl connection $\Ga$
so that $u$ is $\Ga$-geodesic.

For any representative $g\in [g]$ of the conformal structure one can normalize the vector field $u$ to be a $g$-unit, let it be $n^\mu$.

A Weyl connection for $g$ is a connection $\Ga$ such that the compatibility condition 
\ShowLabel{EPSCompatibility} holds true.
If $\Ga$ is a Weyl connection for $g$ it is a Weyl connection for any other metric conformal to $g$. 

For later convenience, by using compatibility condition \ShowLabel{EPSCompatibility} one also has 
$$
\nab{\Ga}_\al  g_{\mu\nu} =2 A_\al g_{\mu \nu}
\qquad\qquad
\nab{\Ga}_\al  g^{\mu\nu} =-2 A_\al g^{\mu \nu}
\fn$$

\CNote{
In fact, one has
$$
\eqalign{
\nab{\Ga}_\al  g_{\mu\nu} = & \nab{g}_\al  g_{\mu\nu} 
- \(g^{\la\ep} g_{\mu\al} -2 \de^\la_{(\mu}\de^\ep_{\al)}\) A_\ep g_{\la\nu}
- \(g^{\la\ep} g_{\nu\al} -2 \de^\la_{(\nu}\de^\ep_{\al)}\) A_\ep g_{\mu \la} =\cr
=&-  \red{g_{\mu\al} A_\nu}  +\uline{A_\al g_{\mu\nu}}+  \blue{A_\mu g_{\al\nu}}
-  \blue{g_{\nu\al}A_\mu}  +\uline{ A_\al g_{\mu \nu}}+  \red{A_\nu g_{\mu \al}} =2 A_\al g_{\mu \nu}\cr
}
\fn$$
and analogously
$$
\eqalign{
\nab{\Ga}_\al  g^{\mu\nu} = & 
 \nab{g}_\al  g^{\mu\nu} 
+ \(g^{\mu\ep} g_{\la\al} -2 \de^\mu_{(\la}\de^\ep_{\al)}\) A_\ep g^{\la\nu}
+ \(g^{\nu\ep} g_{\la\al} -2 \de^\nu_{(\la}\de^\ep_{\al)}\) A_\ep g^{\mu \la} =\cr
=&
 \red{A^\mu \de^\nu_\al} - \uline{A_\al g^{\mu\nu}}- \blue{\de^\mu_{\al} A^\nu}
+  \blue{ A^\nu \de^\mu_\al}  -  \uline{A_\al g^{\mu \nu}}- \red{\de^\nu_{\al} A^\mu} =-2 A_\al g^{\mu \nu}\cr
}
\fn$$
}

We first want to show that there exists in $M$ a Weyl connection such that the integral curves of $n$ are $\Ga$-geodesics.
For that to be true one should determine $\Ga$ so that
$$
n^\mu \nab{\Ga}_\mu n^\nu=\vp n^\nu
\fl{GeodesicEq}$$
holds for some scalar field $\vp$.

The equation \ShowLabel{GeodesicEq} can be expanded by using \ShowLabel{EPSCompatibility}. One finds:
$$
n^\mu \nab{g}_\mu n^\nu  + n^\mu  n^\la\(g^{\nu\ep} g_{\mu\la} -2 \de^\nu_{(\mu}\de^\ep_{\la)}\) A_\ep=\vp n^\nu
\fl{GeoEq2}$$
which should be solved for $A$.

Equation \ShowLabel{GeoEq2} can be recasted under the following form
$$
 A^\nu +2 n^\nu A_\ep n^\ep= n^\mu \nab{g}_\mu n^\nu  -\vp n^\nu
\fl{GeoEq3}$$
and contracted with $n_\nu$ to obtain
$$
\eqalign{
&  A_\ep n^\ep -2A_\ep n^\ep = n^\mu n_\nu \nab{g}_\mu n^\nu +\vp\cr
& A_\ep n^\ep = - \(\frac[1/2]n^\mu  \nab{\ast}_\mu |n|^2 +\vp \)
\equiv    -\vp \cr
}
\fn$$

\CNote{
The following identity 
$$
n_\nu \nab{g}_\mu n^\nu= n^\nu \nab{g}_\mu n_\nu =  \nab{\ast}_\mu |n|^2 -  n_\nu \nab{g}_\mu n^\nu 
\quad \then
n_\nu \nab{g}_\mu n^\nu=\frac[1/2] \nab{\ast}_\mu |n|^2\equiv 0
\fn$$
has been used.
}

This can be  replaced back into equation \ShowLabel{GeoEq3} above to find finally
$$
   A_\nu = n^\mu \nab{g}_\mu n_\nu  + \vp n_\nu
\fl{WeylCovector}$$
Notice how $A$, and hence $\Ga$, is uniquely determined by $g$, $n$ and $\vp$. 
As a consequence there is a family of connections for which the congruence generated by $n$ is $\Ga$-geodesic, 
one for any parametrization of the integral curves of $n$ (which for simplicity and physical reasons we here assume to be complete).
Notice  also that if $n$ happens to be a geodesic field for $\Ga=\{g\}$ then $A=0$, as expected.
The covector $A$ in fact measures the failure of $n$ being $g$-geodesic.

\NewSection{Conformal Invariance}

In Section 2 we started by choosing a representative $g$ of the conformal structure.
Here we want to investigate what would have happened if we had chosen another representative $\tilde g= \Phi^2 g$.

First of all the Weyl connection can be expressed in terms of $\tilde g$ and another covector $\tilde A$ 
satisfying \ShowLabel{EPSCompatibility} for $(\tilde g, \Ga)$ as well.
In fact,
$$
\Ga^\al_{\mu\nu}= \{\tilde g\}^\al_{\mu\nu} + \(\tilde g^{\al\ep} \tilde g_{\mu\nu} -2 \de^\al_{(\mu}\de^\ep_{\nu)}\) \tilde A_\ep=
\{g\}^\al_{\mu\nu} + \(g^{\al\ep} g_{\mu\nu} -2 \de^\al_{(\mu}\de^\ep_{\nu)}\) \(\tilde A_\ep - \del_\ep \ln \Phi\)
\fn$$
Hence we see that the covector \ShowLabel{EPSCompatibility} in transforms as follows
$$
\tilde A := A +  d \ln \Phi
\fl{CTCovector}$$

Also the unit vector $n$ is affected by the conformal transformation. If $n$ is the $g$-unit, when another conformal representative $\tilde g$
is selected one has 
$$
\tilde n^\la=\frac[1/\Phi] n^\la
\fn$$
Analogously, for its covariant version one has
$$
\tilde n_\al = \tilde g_{\al\be} \tilde n^\be= \Phi^2 g_{\al\be} \frac[1/\Phi] n^\be= \Phi n_\al
\fn$$

The geodesic equation \ShowLabel{GeodesicEq} is also affected by the conformal transformation 
$$
\tilde n^\mu \nab{\Ga}_\mu \tilde n^\nu= 
\frac[1/\Phi^2]  n^\mu \nab{\Ga}_\mu n^\nu +  \frac[1/\Phi]  n^\mu \nab{\ast}_\mu \frac[1/\Phi]  n^\nu= 
\frac[1/\Phi]  \( \vp  -   n^\mu \nab{\ast}_\mu \ln\Phi  \) \tilde n^\nu = 
\tilde \vp \tilde n^\nu
\fn$$
where we set $\tilde \vp:=\frac[1/\Phi]  \( \vp  -   n^\mu \nab{\ast}_\mu \ln\Phi  \)$.
Here we denote by $\nab{\ast}_\mu$ the covariant derivative when it happens to be independent of the connection,
i.e.~$\nab{\ast}_\mu=\del_\mu$.

In other words one recovers the conformal invariance by normalizing the vector field and rescaling the metric and the scalar factor
accordingly.

Also equation \ShowLabel{GeoEq2} is preserved by the conformal transformation.
If one started from a conformally equivalent framework, then the equation
$$
\tilde n^\mu \nab{\tilde g}_\mu \tilde n^\nu  + \tilde n^\mu \tilde n^\la \(\tilde g^{\nu\ep} \tilde g_{\mu\la} -2 \de^\nu_{(\mu}\de^\ep_{\la)}\) \tilde A_\ep 
=\tilde \vp \tilde n^\nu
\fn$$
would be obtained and  shown to be equivalent to \ShowLabel{GeoEq2}. 

\CNote{
In fact, one has
$$
\eqalign{
&\tilde n^\mu \nab{\tilde g}_\mu \tilde n^\nu  + \tilde n^\mu \tilde n^\la \(\tilde g^{\nu\ep} \tilde g_{\mu\la} -2 \de^\nu_{(\mu}\de^\ep_{\la)}\) \tilde A_\ep 
=\tilde \vp \tilde n^\nu\cr
&\tilde n^\mu \nab{g}_\mu \tilde n^\nu  + \tilde n^\mu \tilde n^\la \(g^{\nu\ep} g_{\mu\la} -2 \de^\nu_{(\mu}\de^\ep_{\la)}\) \(A_\ep + \red{ \del_\ep \ln \Phi} -\red{\del_\ep \ln \Phi}\)
=\frac[1/\Phi]  \( \vp  -   n^\mu \nab{\ast}_\mu \ln\Phi  \) \tilde n^\nu\cr
& \frac[1/\Phi^2] n^\mu \nab{g}_\mu n^\nu  - \red{ \frac[1/\Phi^2] n^\mu n^\nu \nab{\ast}_\mu\ln \Phi   }
+\frac[1/\Phi^2]   n^\mu n^\la \(g^{\nu\ep} g_{\mu\la} -2 \de^\nu_{(\mu}\de^\ep_{\la)}\)A_\ep
=\frac[1/\Phi^2]  \( \vp  -   \red{n^\mu \nab{\ast}_\mu \ln\Phi}  \) n^\nu\cr
& n^\mu \nab{g}_\mu n^\nu 
+  n^\mu n^\la \(g^{\nu\ep} g_{\mu\la} -2 \de^\nu_{(\mu}\de^\ep_{\la)}\)A_\ep
= \vp  n^\nu\cr
}
$$
}

Finally, the connection determined by the congruence given by equation \ShowLabel{WeylCovector}
is independent of the conformal rescaling.
In fact, one has
$$
\eqalign{
\tilde A_\nu=&  \tilde n^\mu \nab{\tilde g}_\mu \tilde n_\nu  + \tilde \vp \tilde n_\nu
=\tilde n^\mu \nab{g}_\mu \tilde n_\nu + \tilde n^\mu \tilde n_\la \(g^{\la\ep} g_{\nu\mu} -2 \de^\la_{(\mu}\de^\ep_{\nu)}\) \del_\ep \ln \Phi + \tilde \vp \tilde n_\nu
=\cr
=& n^\mu \nab{g}_\mu n_\nu 
+{ n^\mu n_\nu \nab{\ast}_\mu \ln \Phi}  + \( {n_\nu n^\ep}+ \de^\ep_{\nu} -{ n^\ep n_\nu}  \) \del_\ep \ln \Phi 
+  \( \vp  -  {n^\mu \nab{\ast}_\mu \ln\Phi}  \)  n_\nu
=\cr
=& n^\mu \nab{g}_\mu n_\nu 
 + \del_\nu \ln \Phi 
+   \vp   n_\nu
=  A_\nu + \del_\nu \ln \Phi \cr
}
\fn$$
which shows that the covector $\tilde A$ is exactly the covector needed to define the same connection $\Ga$ in the conformal framework,
 as one can easily check by using \ShowLabel{CTCovector}.

\NewSection{Fluids and Geodesics}

In Section $2$ we determined the connection $\Ga$ so that the integral curves of the $[g]$-timelike
vector $u$ are $\Ga$-geodesics.
If one assumes $u$ to generate the flow lines of a fluid then the assumption for them being geodesics should be motivated.
Let us briefly review what happens in standard GR (where the connection $\Ga$ is identified with the Levi-Civita connection
of the metric $g$).

In the case of {\it pure dust}  (i.e.~when there is no pressure, $p=0$) then there is no much doubt; the dust particles
do not interact with anything else than the gravitational field and hence they are freely falling.
In this case, the field can be shown to be geodesics as a consequence of the conservation of the energy-momentum tensor of the fluid;
see \ref{Bibliopolis}.

In fact, in general the energy-momentum tensor of a fluid can be written as
$$
T_{\mu\nu}= p g_{\mu\nu}+ (p+\rho) n_\mu n_\nu
\fl{EMt}$$ 
where $\rho$ represents the density and $p$ represents the pressure  of the fluid. 
The conservation of the energy-momentum tensor is $\na_\mu T^{\mu\nu}=0$.
One can take the scalar product along $n$ to get
$$
\eqalign{
n^\rho g_{\rho\nu} \na_\mu T^{\mu\nu}=&
{\na_\mu p n^\mu}- {n^\mu \na_\mu}({p}+\rho)- (p+\rho)  \na_\mu n^\mu +{(p+\rho) n_\nu n^\mu \na_\mu n^\nu}=\cr
=& - p \na_\mu n^\mu- \na_\mu\(\rho n^\mu\) =0
}
\fn$$ 
where we used the fact that $n$ is a unit vector.
This information can be plugged back into the original conservation law and then
$$
\eqalign{
 \na_\mu T^{\mu\nu}=&
\na_\mu p g^{\mu\nu}+ \( \na_\mu p n^\mu -{p \na_\mu n^\mu} +{p  \na_\mu n^\mu}\) n^\nu+ (p+\rho) n^\mu \na_\mu n^\nu
 =\cr
=& \na_\mu p \( g^{\mu\nu}+  n^\mu  n^\nu\)+ (p+\rho) n^\mu \na_\mu n^\nu =0
}
\fn$$ 
One can easily see that if $p=0$ (and $\rho>0$) then the conservation of energy momentum tensor implies 
$$
n^\mu \na_\mu n^\nu =0
\fn$$
i.e.~$n$ is a geodesics field. If $p\not=0$ the same cannot follow in general along the same lines.

However, one can guess that there might exist special combinations of the pressure gradient and the metric tensor such that
one still has $\na_\mu p \( g^{\mu\nu}+  n^\mu  n^\nu\)=0$; in these cases the vector $n$ is geodesic even if $p\not=0$.
We shall see that Friedman-Robertson-Walker relativistics cosmologies do provide one of these cases.

In Cosmology, the cosmological principle claims that the metric is homogeneous and isotropic; see \ref{catalogo}.
For such metrics there exists a coordinate system in which the metric tensor $g$ is in the form
$$
g= -dt^2 + a^2(t) \( \Frac[dr^2/ 1-kr^2] + r^2 \(d\te^2+ \sin^2\te d\phi^2\)\)
\fl{FRWmetric}$$ 
where the function $a(t)$ is called the {\it scale factor}, and $k=0, \pm 1$.
One should stress that this ansatz is done before fixing any dynamics. 
In fact, any metric in this form is a solution of Einstein equations if one defines the energy-momentum tensor
of matter sources as
$$
G_{\mu\nu} =: \ka T_{\mu\nu}
\fn$$ 
where $G_{\mu\nu}:=R_{\mu\nu}-\frac[1/2]Rg_{\mu\nu}$ is the Einstein tensor of $g$.

One can easily see that the Einstein tensor of a metric in the form \ShowLabel{FRWmetric} is in the form \ShowLabel{EMt},
for a specific choice of pressure $p$ and density $\rho$ as a function of the scale factor $a(t)$ (together with its first and second derivatives) and for $n:= \del_t$. Here the relevant point is that the vector $n$ appearing in the energy momentum tensor  \ShowLabel{EMt} is fixed uniquely by the metric ansatz \ShowLabel{FRWmetric}.

Hence for any spacetime in Cosmology the metric is associated to a fluid matter source and the fluid flow lines are generated by 
 $n= \del_t$. In this case one can check directly that, even when pressure is non-zero, the vector $n$ is geodesic
 for all metrics in the form \ShowLabel{FRWmetric}.

Hence, at least for all dust matter models and all models in Cosmology (even with pressure)
it is reasonable to assume that the flow lines of a fluid are generated by a timelike geodesic congruence.

\NewSection{Conclusions and Perspectives}

We have shown that in Weyl geometries one can select the connection $\Ga$ so that any {\it generic} timelike congruence
is $\Ga$-geodesic.
In EPS framework (and in particular in $f(R)$ models) one can use any $[g]$-timelike congruence 
as the flowlines of a fluid.

The connection $\Ga$ determined by a timelike congruence is conformally invariant.

This research is part of a much larger project aiming to model a generic self-gravitating fluid in EPS formalism.
One can specify the conformal structure to be the Minkowskian one by setting $g=\eta$ and still be free to
model any congruence of (timelike) worldlines as the flow of a fluid. In this framework the gravitational field is encoded into the covector $A$
which in turn determines the Weyl connection $\Ga$. 

Let us stress that in EPS setting there is no freedom in choosing the connection associated to the free fall. In EPS framework the free fall of particles is {\it by construction} described by the connection $\Ga$ while
the Levi-Civita connection of $g$ plays just the kinematical role of {\it reference frame} in the affine space of connections (which is moreover conformally covariant since one can start from any representative of the conformal structure).

The extra degrees of freedom to determine $\Ga$ are thence encoded into the covector $A$ which is kinematically free to be generic.
The dynamics of the theory determines then the connection $\Ga$ fixing the covector $A$ in terms of matter fields and $g$.

Of course Weyl geometries are affected by physical interpretation problems mainly related to the (possibly non-trivial) holonomy of the connection $\Ga$; see \ref{EPS}. 
However, these problems arise only if $\Ga$ is not metric while metric connections do not generate any physical problem of this kind and have just to be interpreted correctly. We stress that in all $f(R)$ models the connection $\Ga$ turns out to be automatically metric, unless one introduces a matter Lagrangian in which matter couples directly with the connection
$\Ga$.  
 
In a forthcoming paper we shall investigate in detail the energy-momentum tensor in a Weyl framework together with its conservation.
In Weyl geometries one has two connections ($\Ga$ and $\{g\}$), not just one as in standard GR.
All derivations and theorems of standard GR need to be generalized to Weyl geometry at least by specifying which is the 
relevant connection to be used.
This has to do also with determining which is the {\it physical} frame in $f(R)$ models; see \ref{Magnano}.
 
\

\Acknowledgements
We acknowledge the contribution of INFN (Iniziativa Specifica NA12) and the local research project 
{\it Leggi di conservazione in teorie della gravitazione classiche e quantistiche} (2010) of Dipartimento di Matematica of University of Torino (Italy).

We wish to thank N.Fornengo for discussions about geodesics and fluids.
We also wish to thank M.Ferraris and G.Magnano for discussions and comments.

\ShowBiblio

\end